\newcommand\aj{{AJ\,}}%
\newcommand\apj{{ApJ\,}}%
\newcommand\apjl{{ApJ\,}}%
\newcommand\apss{{Ap\&SS\,}}%
\newcommand\aap{{A\&A\,}}%
\newcommand\mnras{{MNRAS\,}}%
\newcommand\nat{{Nature\,}}%
\newcommand\solphys{Sol.~Phys.\,}%

\documentclass[graybox, natbib, footinfo]{svmult}

\usepackage{mathptmx}       
\usepackage{helvet}         
\usepackage{courier}        
\usepackage{type1cm}        
\usepackage{makeidx}         
\usepackage{graphicx}        
\usepackage{multicol}        
\usepackage[bottom]{footmisc}

\makeindex             

\begin{document}
\title*{Effects of rotation on stellar evolution and asteroseismology of red giants}
\author{Patrick Eggenberger}
\institute{P. Eggenberger \at Observatoire de Gen\`eve, Universit\'e de Gen\`eve, 51 ch. des Maillettes,
CH-1290 Sauverny\\ \email{patrick.eggenberger@unige.ch}
}
%
%
\maketitle

\abstract{
The impact of rotation on the properties of low-mass stars at different evolutionary stages is first described by discussing the properties of stellar models computed with shellular rotation. The observational constraints that are currently available to progress in our understanding of these dynamical processes are then presented, with a peculiar focus on asteroseismic measurements of red-giant stars.}

\section{Effects of rotation on stellar properties}
\label{sec_dhr}

Rotation can have an important impact on stellar physics and evolution \citep{mae09}. These rotational effects are briefly discussed for different evolutionary phases of low-mass stars by comparing models computed with and without the inclusion of shellular rotation \citep{zah92} with the Geneva stellar evolution code \citep{egg08}. Two main kind of rotational effects can be distinguished: the direct effects on the stellar structure resulting from hydrostatic corrections due to the centrifugal force and the indirect effects due to rotational mixing.

\subsection{Pre-main sequence evolution}

During the pre-main sequence (PMS) evolution of a rotating solar-type star, only hydrostatic effects of rotation have a significant impact on the global stellar properties. As a result of the decrease of the effective gravity when rotation is taken into account, the evolutionary track in the HR diagram of a rotating PMS model is slightly shifted to lower effective temperatures and luminosities; it is then similar to the track of a non-rotating star computed with a slightly lower initial mass. This shift observed in the HR diagram is directly related to the surface velocity of the model and is thus more pronounced for fast rotators. 

Rotational mixing does not significantly influence the global stellar parameters during the PMS, but can change the surface abundances of light elements. Rotation is indeed found to increase the lithium depletion during the PMS by generating turbulence below the convective envelope through the shear instability. This decrease in the surface lithium abundance by rotational mixing is more directly related to the increase of differential rotation in the stellar interior than to the value of the rotational velocity at the stellar surface. The degree of differential rotation in the radiative zone of a PMS rotating model is sensitive to the duration of the disc-locking phase, during which the surface velocity of the model is simply assumed to remain constant. Consequently, stellar models computed with longer disc lifetimes exhibit lower surface lithium abundances on the zero-age main sequence (ZAMS) than models that experience a shorter disc-locking phase. Moreover, a model computed with a longer duration of the disc-locking phase also loses a larger amount of angular momentum and reaches the ZAMS with a lower surface rotational velocity. An interesting relation between the surface velocity and the lithium content on the ZAMS is then obtained, since stars with lower rotation rates on the ZAMS are predicted to be more depleted in lithium than stars that are fast rotators on the ZAMS \citep[see][for more details]{egg12_pms}. Such a relation seems to be in good agreement with observations of lithium abundances and surface rotation rates in the Pleiades \citep[e.g.][]{sod93}.


\subsection{Main-sequence evolution}

During the main-sequence (MS) evolution of a solar-type star, hydrostatic effects of rotation are only observed near the ZAMS, where rotating models are characterized by slightly lower luminosities and effective temperatures than non-rotating ones. Due to the magnetic braking undergone by solar-type stars, the surface rotational velocities of these models rapidly decrease during the MS and the impact of the centrifugal force on the global stellar parameters rapidly becomes negligeable. The effects of rotational mixing on the global stellar parameters become however more and more important during the evolution on the MS. Larger effective temperatures and luminosities are then found for rotating models compared to non-rotating ones. For models of solar-type stars characterized by a radiative core, this is mainly due to the fact that rotational mixing counterbalances atomic diffusion in the external layers of the star \citep{egg10_sl}. This results in higher values of helium abundance at the surface of rotating models and hence to a decrease of the opacity in the external layers of the star, which explains the shift to the blue part of the HR diagram when rotational effects are taken into account. For more massive models with a convective core during the MS, the inclusion of rotational effects increases the size of the core and changes the chemical profiles in the radiative zone. This results in an increase of the luminosity and a widening of the MS for rotating models compared to non-rotating ones \citep[e.g.][]{egg10_rg}. Rotational mixing also changes the properties in the central stellar layers by transporting fresh hydrogen fuel into the stellar core. This increases the value of the central abundance of hydrogen at a given age and leads to an enhancement of the main-sequence lifetime for rotating models compared to non-rotating ones.

The changes of the global stellar parameters and of the properties of the central layers induced by rotation have also an impact on the asteroseismic properties of solar-type stars \citep{egg10_sl}. As a result of the decrease of the stellar radius when rotational effects are taken into account, the value of the mean large frequency separation at a given age is lower for rotating models than for non-rotating ones. Rotating models are also characterized by higher values of the mean small separation at a given age than non-rotating ones, because of the transport of hydrogen in the central layers by rotational mixing.

\subsection{Post-main sequence evolution}

Although red giants are generally characterized by low surface rotational velocities, the impact of rotation on these stars can be significant, because the evolution in the red giant phase is sensitive to the rotational history of the star and in particular to the changes of the stellar properties during the MS. For models massive enough to ignite helium burning in non-degenerate conditions, the increase of the luminosity when rotational effects are taken into account leads to a shift to higher luminosities of the location of the core helium-burning phase for rotating models compared to non-rotating ones. For low-mass red-giant models that undergo the helium flash, the situation is different, since the location of the core helium burning phase in the HR diagram is very similar for rotating and non-rotating models. For these stars, the inclusion of rotation has an impact on the location of the bump; at solar metallicity, rotating models exhibit a lower luminosity and a higher effective temperature at the bump than non-rotating ones \citep[e.g.][]{cha10,egg12_procrg}.

\section{Asteroseismic studies of red-giant stars}
\label{sec:astero_rot}

The brief description of the effects of rotation on the properties of low-mass stars presented above has been done in the context of shellular rotation. The outputs of these rotating models, and in particular the quantitative impact of rotational mixing, are sensitive to the prescriptions used for the modelling of meridional circulation and shear instabilities \citep{mey13}. It is thus interesting to consider the observational constraints that are currently available to constrain these dynamical processes, with a peculiar focus on asteroseismic data. 

The internal rotation profile of the Sun deduced from helioseismic measurements \citep{bro89,els95,kos97,cou03,gar07} is a key observational constraint for rotating models. These observations show a nearly flat rotation profile, while solar models including only shellular rotation predict a rapidly rotating core \citep{pin89,cha95,egg05_mag,tur10}. This is a clear indication that another physical process is at work in the solar radiative interior. This still undetermined mechanism could be related to magnetic fields \citep[e.g.][]{cha93,gou98,egg05_mag} or to angular momentum transport by internal gravity waves \citep[e.g.][]{zah97,tal02,cha05,mat13}. The helioseismic results stimulated various attempts to detect and characterize solar-like oscillations for other stars than the Sun. Ground-based asteroseismic observations have enabled the detection of solar-like oscillations for a limited number of bright solar-type stars \citep[see e.g.][]{bed08} and for a few red giants \citep{fra02,bar04,der06}. Thanks to the CoRoT \citep{bag06} and {\it Kepler} \citep{bor10} space missions, the observation and characterization of solar-like oscillations have now been obtained for a very large number of stars. 

Concerning the rotational properties of stars, these seismic observations have led to the determination of rotational frequency splittings of mixed modes in red giants \citep{bec12,deh12,mos12}. By lifting the degeneracy in the azimuthal order of non-radial modes, rotation leads to $(2\ell+1)$ frequency peaks in the power spectrum for each mode. Rotational splittings are then defined as the frequency separations between these peaks. These frequency spacings are related to the angular velocity and the properties of the modes in their propagation regions. Mixed modes are sensitive to the properties in stellar interiors and contain a different amount of pressure and gravity-mode influence. An oscillation mode dominated by its acoustic character will be more sensitive to the properties in the external layers of a red giant, while a gravity-dominated mode will be sensitive to the properties in the stellar core. Rotational splittings of mixed modes thus contain valuable information about internal rotation of red giants, which are of prime importance to better understand the dynamical physical processes at work in stellar interiors \cite[see e.g.][]{egg12_rg,mar13,gou13}.

\subsection{Mixed modes in the red giant KIC~8366239}

To illustrate in more details how the observation of rotational splittings of mixed modes in red giants can constrain the modelling of internal angular momentum transport, we first consider the case of the red giant KIC~8366239 observed by the {\it Kepler} spacecraft \citep{bec12}. The precise measurements of rotational splittings in KIC~8366239 show radial differential rotation in the stellar interior with central layers rotating at least ten times faster than the envelope \citep[see][for more details]{bec12}.

By performing the modelling of this asteroseismic target, a model in the H-shell burning phase with a solar chemical composition and an initial mass of 1.5\,M$_{\odot}$ is determined. This star is thus found to exhibit a convective core during its evolution on the main sequence. The theoretical rotational splittings corresponding to rotating models of KIC~8366239 are then computed and compared to the observed values. These rotating models predict an increase of the angular velocity in the stellar core, which seems at first sight to be in good agreement with the asteroseismic observations. However, the values of rotational splittings predicted by models including shellular rotation only are found to be much larger than the observed values, even in the case of models computed for very low initial velocities on the ZAMS \citep[see][for more details]{egg12_rg}. This discrepancy is mainly due the large increase in the rotational velocity predicted in the core of rotating models. The internal rotation profile of models including shellular rotation is indeed found to be very steep during the red giant phase as a result of the central contraction occurring at the end of the main-sequence evolution \citep[e.g.][]{pal06,egg10_rg,egg12_rg,mar13}. We then conclude that meridional circulation and shear instability alone produce an insufficient coupling to correctly reproduce the observed values of rotational splittings in KIC~8366239. This shows that an additional mechanism for the internal transport of angular momentun is at work during the post-main sequence evolution in order to obtain a predicted rotation profile during the red giant phase in agreement with asteroseismic data.

The observed values of rotational splittings for mixed modes in KIC~8366239 can be used to place constraints on the efficiency of this unknown additional mechanism for the internal transport of angular momentum. For this purpose, rotating models of the red giant KIC~8366239 are computed by adding a viscosity term corresponding to this unknown additional physical process in the equation describing the transport of angular momentum by the meridional circulation and the shear instability. In a first step, this viscosity is simply assumed to be constant. The rotation profiles predicted by models computed with different values for this additional viscosity are then confronted to asteroseismic data to determine the efficiency needed for this transport process to correctly reproduce the observed splittings. We then found that the ratio of rotational splittings for dipole gravity-dominated modes to those for modes dominated by their acoustic character strongly constrains the efficiency of this additional transport mechanism of angular momentum. In the case of the red giant KIC~8366239, a viscosity of $3 \times 10^{4}$\,cm$^2$\,s$^{-1}$ is determined for this unknown additional process \citep{egg12_rg}. Interestingly, this value deduced from asteroseismic observations of a red giant is quite similar to the value required to correctly reproduce the spin-down of slowly rotating solar-type stars observed during the beginning of the evolution on the main sequence \citep{den10_spin}. This may suggest that the same unknown transport mechanism of angular momentum is at work during the main-sequence and the post-main sequence evolution.

\subsection{Mixed modes in the red giant KIC~7341231}

Rotational splittings have also been precisely characterized for 18 oscillation modes in the red giant KIC~7341231 observed during one year by the {\it Kepler} space mission \citep{deh12}. These splittings have been used to perform an inversion of the rotation profile showing radial differential rotation with a rotation rate in the central stellar layers at least five times higher than in the envelope (see \cite{deh12} for more details). This red giant star is characterized by a low metallicity ([Fe/H]$ \approx -1$) and a low mass of about 0.84\,M$_\odot$. Consequently, this star exhibits a radiative core during the main sequence. It is thus particularly interesting to compare the rotational properties of this target with the results describe above for KIC~8366239, which is a more massive star with a convective core during the main sequence.

Rotation profiles predicted by models of KIC~7341231 including a detailed treatment of shellular rotation have been compared to the internal rotation profile deduced form the asteroseimic measurements of KIC~7341231 \cite{cei12, cei13}. Such a comparison indicates that rotating models including shellular rotation only lead to internal rotation profiles that are too steep compared to the observed one. This shows that the efficiency of angular momentum transport through meridional circulation and shear instability is not sufficient to correctly reproduce the asteroseismic data obtained for a low-mass red giant like KIC~7341231. An additional mechanism for the internal transport of angular momentum is thus also needed in this case. The studies of KIC~8366239 and KIC~7341231 thus suggest that this unknown physical process is at work in the interiors of red giant stars exhibiting a convective core as well as a radiative core during the evolution on the main sequence. In the case of the low-mass star KIC~7341231, it is interesting to note that the efficieny of the internal transport of angular momentum during the main sequence has an important impact on the rotation profiles predicted during the red giant phase \citep[see][for more details]{cei13}. \cite{tay13} obtained similar results from limiting case scenarios of solid-body rotation or local conservation of angular momentum in radiative zones during the main-sequence and post-main-sequence evolution.

\subsection{Rotational splittings for a large sample of red giants}

In addition to the detailed study of internal rotation profiles for a limited number of red giants, rotational frequency splittings have been obtained for a large sample of about 300 red-giant stars observed with the {\it Kepler} spacecraft \citep{mos12}. This study shows that radial differential rotation is present for all red giants of the sample and provides an estimate of the mean core rotation period for red giant stars at different evolutionary stages. Interestingly, the mean core rotation period is found to be larger for red clump stars than for red-giant branch stars, suggesting that an efficient mechanism for the internal transport of angular momentum is at work during the red-giant phase to explain the observed spin-down of the stellar core.

The comparison between these estimates of core rotation rates for red giants at different evolutionary stages and theoretical predictions of rotating models offers a particularly valuable mean to progress in the modelling of the different mechanisms responsible for the angular momentum transport in stellar interiors. A first comparison has been recently performed between the core rotation periods deduced from asteroseismic data for evolved secondary-clump stars and rotating models assuming solid-body rotation or local conservation of angular momentum in radiative zones \citep{tay13}. This study indicates that models computed by assuming solid-body rotation during the whole stellar evolution are able to correctly reproduce the observed core rotation rates of secondary-clump stars. By assuming solid-body rotation, these models are at odds with the observations of radial differential rotation in red giants, but they clearly underline the need of an efficient mechanism for angular momentum transport to correctly account for the asteorseismic data obtained for these stars.

\section{Conclusion}

We have first briefly described the impact of shellular rotation on the properties of low-mass stars at different evolutionary stages. Including rotational effects leads to changes of the global and asteroseismic properties of these stars. It will be particularly interesting to investigate in more details the impact of such changes on the global properties of stars in clusters \citep[e.g.][]{gir11} and on population studies of red giants observed by current asteroseismic space missions \citep[as done for instance by][using non-rotating models]{mig09a}.
The comparison between models computed with shellular rotation and asteroseismic measurements of red giants suggests that an additional mechanism for the internal transport of angular momentun is at work during the post-main sequence evolution. This still undetermined process could be related to the transport of angular momentum by internal gravity waves \citep[e.g.][]{cha05} or magnetic fields \citep[e.g.][]{egg10_magn}. This illustrates that we are still far from a satisfying description of the transport processes at work in stellar interiors and that asteroseismic measurements can help us progress in the modelling of these mechanisms.

\begin{acknowledgement}
Part of this work was supported by the Swiss National Science Foundation.
\end{acknowledgement}
%


\end{document}